\documentclass{article}
\usepackage{spconf,amsmath,graphicx,hyperref,booktabs}
\usepackage{multirow}
\usepackage{relsize}
\usepackage{xcolor}
\usepackage[table]{xcolor}
\usepackage{cite}

\title{When Capabilities Fail to Compose: Diagnosing the Compositionality Gap in Large Audio-Language Models}
\name{
\begin{tabular}{c}
   Chien-Feng Liu$^{1,3}$, Chih-Kai Yang$^{1*}$, Bo-Han Feng$^{1*}$, Yu-Hsuan Li Liang$^{1*}$\thanks{*Equal Contribution.} \\ 
   Hung-yi Lee$^{1,2}$, Cheng-Fu Chou$^{1}$
\end{tabular}
}
\address{$^{1}$National Taiwan University \\
$^{2}$NTU Artificial Intelligence Center of Research Excellence (NTU AI-CoRE) \\
$^{3}$ASUS Open Cloud Infrastructure Software Center}
\begin{document}
\ninept
\maketitle
\begin{abstract}

Large audio-language models (LALMs) perform strongly on individual audio tasks, but whether these capabilities can be reliably composed remains underexplored. We conduct a controlled diagnostic study of capability composition in LALMs, requiring models to integrate audio-attribute recognition, cue-conditioned segment selection, and downstream ASR or question answering. We construct two-utterance inputs with distinct acoustic cues to evaluate composition across environmental sound, gender, and emotion cues, with ASR, Math QA, and Factual QA as downstream tasks. Across four open-source LALMs, compositional QA accuracy decreases in 39 of 40 model–task–cue settings, by an average of 26.7 percentage points. ASR exhibits a similarly consistent degradation, with WER increasing in 39 of 40 settings by an average of 28.5 percentage points, while the magnitude of degradation varies across models, cue types, and cue salience. We further probe these failures through output format, positional preference, and chain-of-thought (CoT) analyses. Our study reveals a systematic gap between possessing individual audio capabilities and reliably composing them. Our code is available at: \href{}{https://github.com/steven-lunar/audio-compositionality-gap}.

\end{abstract}

\section{Introduction}
\label{sec:intro}

\begin{figure}[t]
    \centering
    \includegraphics[width=0.85\linewidth]{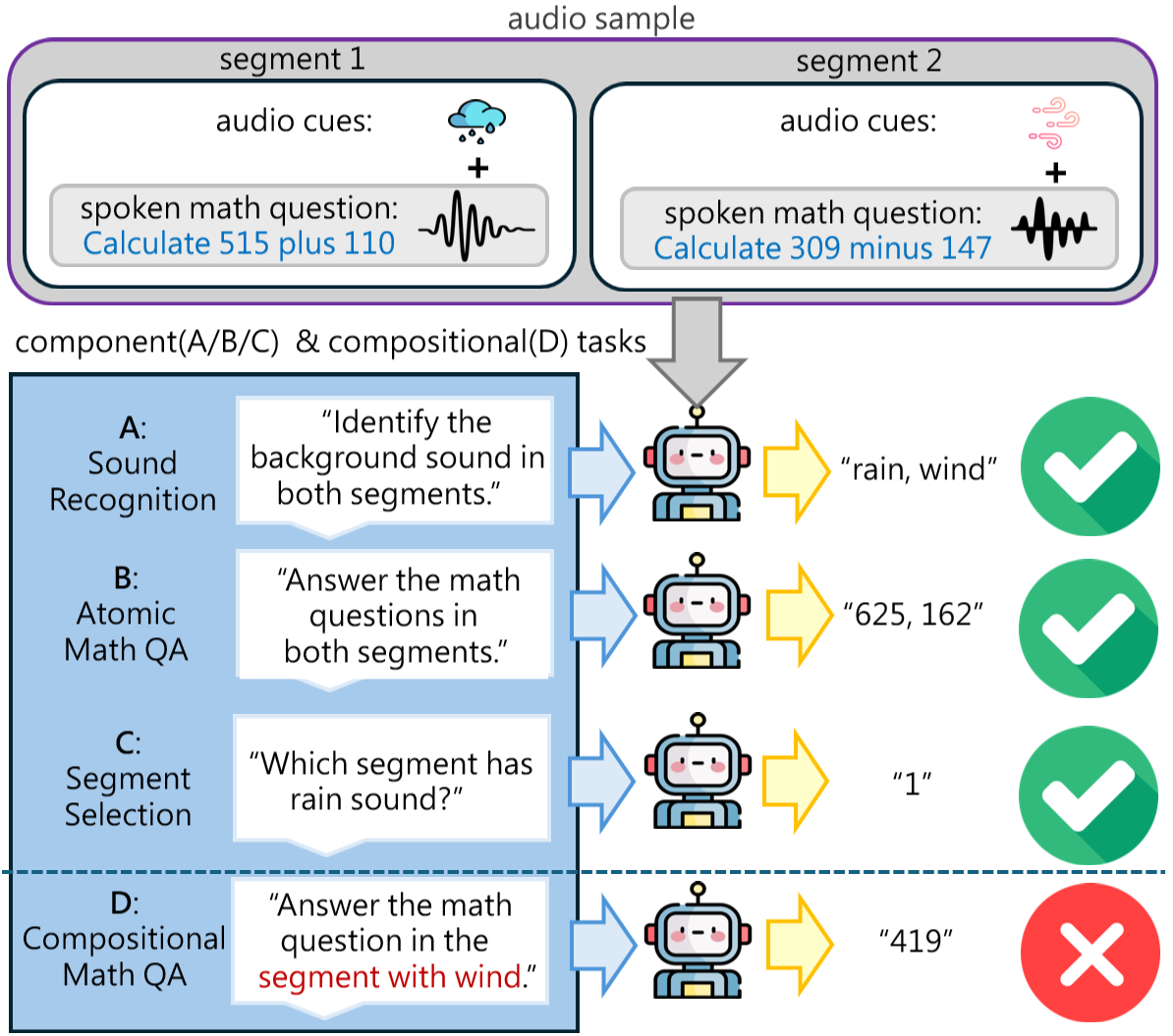}
    \caption{Illustration of our diagnostic setting. Despite succeeding at background sound recognition, atomic math QA, and segment selection, the model fails when these capabilities must be composed to answer the cue-matched question.}
    \label{fig:overview}
    \vspace{-15pt}
\end{figure}

Recent large audio-language models (LALMs)~\cite{Qwen2.5-Omni, minicpm, mimo, kimi} have demonstrated strong capabilities across diverse audio-related tasks in isolation~\cite{mmau, mmaupro, mmar, mmsu}. Yet complex audio understanding often requires integrating multiple capabilities within a single task, e.g., identifying speech associated with a particular acoustic attribute before transcribing or reasoning over its content, as shown in Fig.~\ref{fig:overview}. Prior studies have revealed a \textit{compositionality gap} in large language models (LLMs)~\cite{compgap, compositionalinvest, agentcoma} and vision-language models (VLMs)~\cite{vlmcompgrounding, vlmskillcompgap, vlmcompsurvey}, where models succeed on component skills yet fail when combining these skills. Such failures persist across multi-hop reasoning, heterogeneous skill composition, and visual-skill composition~\cite{compgap, agentcoma, vlmskillcompgap}.

Recent work has begun to investigate reasoning beyond individual audio capabilities~\cite{mugen,sakura,art,compa,polybench,parallm}. SAKURA~\cite{sakura} studies multi-hop reasoning from audio-derived information, while ParA-LLM~\cite{parallm} extends single-attribute understanding to joint reasoning over multiple paralinguistic and acoustic attributes. CompA~\cite{compa} and PolyBench~\cite{polybench} examine compositional structure among audio events, including attribute binding, event order, and concurrent-event reasoning. ART~\cite{art} more directly combines heterogeneous audio capabilities within composite tasks, but does not explicitly compare composite performance with its constituent capabilities. Prior work has provided limited analysis of how individually measured capabilities change under controlled composition and where failures arise in this process.

In this work, we conduct a controlled diagnostic evaluation of capability composition in LALMs, decomposing compositional tasks into component capabilities, cue-conditioned segment selection, and downstream execution. We study three types of acoustic cues (environmental sound, gender, and emotion) and three downstream tasks (ASR, math QA, and factual QA) across four LALMs. We observe substantial degradation from atomic to compositional settings. Strong atomic recognition does not guarantee reliable segment selection, and our analyses indicate that compositional failures cannot be attributed to a single bottleneck.

Our contributions are threefold: (1) we introduce a controlled evaluation framework and dataset for studying capability composition in LALMs; (2) we reveal substantial atomic-to-compositional degradation across models, cue types, and downstream tasks; and (3) we analyze positional and formatting effects and assess the impact of explicit chain-of-thought (CoT) on compositional performance.

\section{Task Formulation and Data Construction}

\subsection{Task Formulation}
\label{sec:definition}



Each sample consists of two sequential speech segments, $X=S_1\Vert S_2$. Each segment $S_i$ is associated with an acoustic attribute $a_i$, drawn from environmental sound, speaker gender, or emotion, and a downstream target $y_i$ corresponding to ASR, math QA, or factual QA. We use $i^*\in\{1,2\}$ to denote the index of the target segment. A textual query $c_{i^*}$ specifies the target attribute $a_{i^*}$, while the other segment serves as a distractor. As illustrated in Fig.~\ref{fig:overview}, the model must use the queried attribute to identify the target segment before performing the downstream task. Table~\ref{tab:tasks} summarizes our evaluation at three levels: component capabilities, cue-conditioned segment selection, and full compositional execution.


\begin{table}[t]
\centering
\caption{Task decomposition. $T$ denotes the downstream task and $i^*$ the target segment index.}
\label{tab:tasks}
\begin{tabular}{lcc}
\toprule
Task & Input & Output \\
\midrule
Atomic recognition  & $X$            & $(a_1,a_2)$ \\
Segment selection & $(X,c_{i^*})$      & $i^*$ \\
Atomic downstream & $(X,T)$        & $(y_1,y_2)$ \\
Compositional      & $(X,c_{i^*},T)$    & $y_{i^*}$ \\
\bottomrule
\end{tabular}
\vspace{-15pt}
\end{table}

Atomic tasks measure the required component capabilities without attribute-conditioned selection: the model predicts the attribute or downstream output for both segments in a single response. Here, \emph{atomic} refers to the absence of cross-skill composition rather than a single-segment input, controlling for input length and multi-segment context. Segment selection isolates attribute-to-segment binding by asking the model to predict the target index $i^*\in\{1,2\}$ from $X$ and $c_{i^*}$. Finally, compositional tasks require the model to identify the segment matching $c_{i^*}$ and execute the downstream task $T$ on that segment. Together, these tasks separate failures in component perception, segment binding, and downstream execution.

\subsection{Dataset Construction}
\label{sec:dataset}

We synthesize speech using CosyVoice 3~\cite{cosyvoice3} with reference audio from CREMA-D~\cite{crema-d}. We select six speakers balanced by gender, each covering three emotions: angry, sad, and happy. Factual QA uses 1,000 questions sampled from TriviaQA~\cite{triviaqa}, while math QA uses Spoken-MQA~\cite{spokenmqa}; the same synthesized utterances are reused for ASR, with the spoken text as the transcription reference.

For environmental-sound cues, we use five ESC-50~\cite{esc50} classes: wind, sea waves, rain, crickets, and chirping birds. Background audio is looped or cropped to the speech duration and mixed at \(0\), \(10\), or \(20\) dB SNR using RMS energy. Each sample concatenates two segments with a 1.5-s silence interval. For the evaluated attribute, \(c_1\neq c_2\), we ensure a unique target. Target attributes, target positions, and distractor classes are balanced, with the target appearing equally often in the first and second positions.

For both math QA and factual QA, we construct four 600-sample subsets: three environmental-sound subsets at \(0\), \(10\), and \(20\) dB SNR, and one subset containing gender- and emotion-conditioned examples, yielding 4,800 two-segment samples in total. We perform automatic quality checks using emotion2vec~\cite{emotion2vec} to verify emotion consistency, UTMOSv2~\cite{utmosv2} to assess speech quality, and Whisper~\cite{whisper} to measure transcription accuracy against the source transcript. All synthesized samples are further manually inspected for intelligibility, attribute consistency, and label correctness, with invalid samples regenerated or removed.

\section{Experimental Setup}
\label{sec:setup}



We evaluate four open-source LALMs: Qwen2.5-Omni-7B~\cite{Qwen2.5-Omni}, MiniCPM-o 4.5~\cite{minicpm}, MiMo-Audio-7B~\cite{mimo}, and Kimi-Audio-7B~\cite{kimi}. All models are evaluated zero-shot with task-specific instructions, XML-style output formats, and greedy decoding on an NVIDIA RTX A6000.



For all atomic tasks, the two segment-level predictions produced in a single response are scored independently rather than jointly at the sample level. Atomic recognition and segment selection are evaluated by accuracy. For downstream tasks at both the atomic and compositional stages, math QA is evaluated by exact match, factual QA by normalized exact match over answer aliases, and ASR by normalized WER. We evaluate these outputs using both tag-based parsing, which extracts predictions from the required fields, and LLM-based parsing with GPT-4o-mini~\cite{gpt4o}, which extracts the intended prediction from the raw output. Comparing the two parsing methods allows us to assess sensitivity to output-format compliance. Under tag-based parsing, unparseable outputs are treated as incorrect for QA and as empty hypotheses for ASR.

\section{Results and Analysis}
\label{sec:results}

\subsection{Main Results}
\label{sec:main_results}

\begin{table}[t]
\centering
\caption{
Atomic recognition and segment selection accuracy (\%) on the math and factual QA subsets. E0/E10/E20 denote environmental-sound conditions at 0/10/20 dB SNR; Gen. and Emo. denote gender and emotion.
}
\label{tab:attr_selection}
\small

\resizebox{0.95\linewidth}{!}{
\begin{tabular}{llccccc}
\toprule
\rowcolor{gray!30} \multicolumn{7}{c}{\textbf{Math QA}} \\
\midrule
\textbf{Model} & \textbf{Task}
& \textbf{E0}
& \textbf{E10}
& \textbf{E20}
& \textbf{Gen.}
& \textbf{Emo.} \\
\midrule

\multirow{2}{*}{Qwen2.5-Omni}
& Recognition & 58.33 & 50.33 & 41.75 & 95.42 & 35.50 \\
& Selection & 87.33 & 84.17 & 75.50 & 76.67 & 72.67 \\
\midrule

\multirow{2}{*}{MiniCPM-o}
& Recognition & 50.25 & 44.50 & 37.33 & 99.42 & 45.50 \\
& Selection & 67.17 & 63.67 & 58.50 & 94.83 & 68.33 \\
\midrule

\multirow{2}{*}{MiMo-Audio}
& Recognition & 42.83 & 38.42 & 30.75 & 83.17 & 40.83 \\
& Selection & 67.67 & 64.33 & 60.50 & 65.83 & 60.50 \\
\midrule

\multirow{2}{*}{Kimi-Audio}
& Recognition & 53.42 & 49.33 & 43.92 & 85.08 & 48.50 \\
& Selection & 79.00 & 74.17 & 67.83 & 71.83 & 66.33 \\
\bottomrule
\end{tabular}
}
\vspace{1.2mm}

\resizebox{0.95\linewidth}{!}{
\begin{tabular}{llccccc}
\toprule
\rowcolor{gray!30} \multicolumn{7}{c}{\textbf{Factual QA}} \\
\midrule
\textbf{Model} & \textbf{Task}
& \textbf{E0}
& \textbf{E10}
& \textbf{E20}
& \textbf{Gen.}
& \textbf{Emo.} \\
\midrule

\multirow{2}{*}{Qwen2.5-Omni}
& Recognition & 59.50 & 50.92 & 42.67 & 98.17 & 35.75 \\
& Selection & 88.67 & 81.67 & 75.67 & 76.17 & 73.00 \\
\midrule

\multirow{2}{*}{MiniCPM-o}
& Recognition & 51.67 & 47.50 & 38.58 & 99.08 & 41.75 \\
& Selection & 72.50 & 67.33 & 63.67 & 91.50 & 63.17 \\
\midrule

\multirow{2}{*}{MiMo-Audio}
& Recognition & 41.08 & 37.00 & 31.42 & 71.50 & 41.92 \\
& Selection & 65.00 & 63.17 & 60.17 & 57.17 & 50.67 \\
\midrule

\multirow{2}{*}{Kimi-Audio}
& Recognition & 52.50 & 52.92 & 45.67 & 85.42 & 49.00 \\
& Selection & 79.67 & 70.67 & 64.00 & 63.83 & 61.83 \\
\bottomrule
\end{tabular}
}
\vspace{-15pt}
\end{table}

\begin{table*}[t]
\centering
\caption{
Atomic and compositional performance. ASR is evaluated by WER (\%, lower is better) and QA by accuracy (\%); each entry reports tag-based / LLM-based parsing results. E0/E10/E20 denote environmental-sound conditions at 0/10/20 dB SNR, and Gen./Emo. denote gender/emotion conditions. Atomic Gen./Emo. values are shared because both attributes use the same audio.
}
\label{tab:main_results}
\vspace{1mm}
\small

\resizebox{\textwidth}{!}{%
\begin{tabular}{llccccc|ccccc}
\toprule
\rowcolor{gray!30} \multicolumn{12}{c}{Math QA} \\
\midrule
&& \multicolumn{5}{c|}{\textbf{ASR ($\downarrow$)}}
& \multicolumn{5}{c}{\textbf{Math QA ($\uparrow$)}} \\
\cmidrule(lr){3-7}
\cmidrule(lr){8-12}

\textbf{Model} & \textbf{Stage}
& \textbf{E0} & \textbf{E10} & \textbf{E20} & \textbf{Gen.} & \textbf{Emo.}
& \textbf{E0} & \textbf{E10} & \textbf{E20} & \textbf{Gen.} & \textbf{Emo.} \\
\midrule

\multirow{2}{*}{Qwen2.5-Omni}
& Atomic
& 9.26/9.31 & 4.29/4.35 & 4.06/4.10 & \multicolumn{2}{c|}{3.86/3.86}
& 70.92/71.33 & 82.83/82.83 & 83.42/83.50 & \multicolumn{2}{c}{82.08/82.42} \\
& Comp.
& 29.05/29.05  & 31.81/31.81  & 42.59/42.59  & 18.62/18.67  & 48.74/48.79 
& 34.33/42.83  & 34.83/47.33  & 28.67/40.50  & 42.00/47.67  & 29.33/40.17 \\
\midrule

\multirow{2}{*}{MiniCPM-o}
& Atomic
& 30.07/15.30 & 21.77/9.82 & 19.72/8.92 & \multicolumn{2}{c|}{18.95/9.19}
& 75.33/75.33 & 86.08/86.17 & 86.83/86.92 & \multicolumn{2}{c}{88.50/88.67} \\
& Comp.
& 67.68/51.01  & 73.53/52.67  & 78.73/57.27  & 58.43/20.16  & 65.69/45.44
& 11.00/58.17  & 13.83/60.17  & 14.83/58.50  & 26.33/93.33  & 15.17/55.33 \\
\midrule

\multirow{2}{*}{MiMo-Audio}
& Atomic
& 13.04/18.23 & 5.68/4.57 & 4.33/3.36 & \multicolumn{2}{c|}{9.56/13.87}
& 60.92/64.83 & 75.08/79.50 & 76.25/81.67 & \multicolumn{2}{c}{77.33/82.83} \\
& Comp.
& 38.08/35.96  & 32.55/30.99  & 37.75/36.14  & 18.58/17.50  & 39.25/37.80 
& 26.67/30.17  & 29.33/35.17  & 26.50/30.50  & 42.17/46.00  & 25.67/27.17  \\
\midrule

\multirow{2}{*}{Kimi-Audio}
& Atomic
& 59.26/42.67 & 47.58/35.78 & 45.60/38.20 & \multicolumn{2}{c|}{42.98/39.68}
& 69.33/70.75 & 80.83/80.92 & 83.42/83.42 & \multicolumn{2}{c}{82.83/82.92} \\
& Comp.
& 64.83/48.48  & 61.10/47.65  & 63.86/54.65  & 83.71/40.27  & 86.69/67.27 
& 18.33/30.50  & 23.17/30.83  & 23.50/30.00  & 11.00/38.33  & 24.50/35.17 \\

\bottomrule
\end{tabular}%
}

\vspace{1.2mm}

\resizebox{\textwidth}{!}{%
\begin{tabular}{llccccc|ccccc}
\midrule
\rowcolor{gray!30} \multicolumn{12}{c}{Factual QA} \\
\midrule
&& \multicolumn{5}{c|}{\textbf{ASR ($\downarrow$)}}
& \multicolumn{5}{c}{\textbf{Factual QA ($\uparrow$)}} \\
\cmidrule(lr){3-7}
\cmidrule(lr){8-12}

\textbf{Model} & \textbf{Stage}
& \textbf{E0} & \textbf{E10} & \textbf{E20} & \textbf{Gen.} & \textbf{Emo.}
& \textbf{E0} & \textbf{E10} & \textbf{E20} & \textbf{Gen.} & \textbf{Emo.} \\
\midrule

\multirow{2}{*}{Qwen2.5-Omni}
& Atomic
& 11.35/12.66 & 5.19/6.46 & 5.06/6.03 & \multicolumn{2}{c|}{5.40/8.28}
& 23.42/23.42 & 29.08/29.08 & 30.83/30.92 & \multicolumn{2}{c}{33.83/33.83} \\
& Comp.
& 30.05/33.50  & 37.02/40.08  & 44.50/47.69  & 17.13/20.69  & 53.04/56.25
& 10.00/10.50  & 12.33/13.17  & 10.50/10.83  & 20.50/20.67  & 12.00/12.00
\\
\midrule

\multirow{2}{*}{MiniCPM-o}
& Atomic
& 27.41/19.79 & 14.72/13.18 & 9.06/8.55 & \multicolumn{2}{c|}{9.20/8.59}
& 30.67/30.67 & 39.58/39.67 & 41.08/41.08 & \multicolumn{2}{c}{41.42/41.42} \\
& Comp.
& 59.72/58.75  & 64.81/63.47  & 69.27/67.97  & 35.99/34.41  & 66.36/62.40
& 14.00/18.17  & 15.17/19.67 & 17.33/23.17  & 25.83/35.33  & 3.83/22.67 \\
\midrule

\multirow{2}{*}{MiMo-Audio}
& Atomic
& 34.00/16.78 & 26.20/9.56 & 27.42/11.71 & \multicolumn{2}{c|}{44.01/25.47}
& 17.50/18.50 & 22.83/24.00 & 25.08/27.17 & \multicolumn{2}{c}{24.83/29.58} \\
& Comp.
& 48.84/47.00  & 50.07/49.78  & 54.46/53.26  & 52.22/55.36  & 70.20/67.39
& 2.83/5.83  & 2.83/6.17  & 1.83/5.83  & 2.83/7.83  & 2.17/5.50 \\
\midrule

\multirow{2}{*}{Kimi-Audio}
& Atomic
& 75.56/65.23 & 74.07/57.15 & 72.13/54.71 & \multicolumn{2}{c|}{61.97/47.38}
& 5.42/25.92 & 10.25/32.67 & 15.33/34.92 & \multicolumn{2}{c}{20.25/37.67} \\
& Comp.
& 64.91/62.56  & 67.52/65.78  & 66.97/65.91  & 98.53/78.65  & 97.60/85.02
& 2.67/17.00  & 5.33/20.67  & 6.00/21.67  & 6.17/22.67  & 6.67/19.67 \\

\bottomrule
\end{tabular}%
}
\vspace{-15pt}
\end{table*}

Table~\ref{tab:attr_selection} reports atomic recognition and segment selection performance. Table~\ref{tab:main_results} compares atomic and compositional downstream performance. Two broad patterns emerge across the evaluated models and conditions. First, compositional performance falls substantially below its atomic counterpart across models, cue types, and downstream tasks. Models that transcribe or answer accurately in the atomic setting often degrade markedly once their outputs must be conditioned on the queried cue. Second, strong atomic recognition does not guarantee accurate segment selection.

\textbf{The relationship between atomic recognition and segment selection is cue-dependent, and strong recognition does not necessarily imply reliable selection.} Table~\ref{tab:attr_selection} reports the accuracy of atomic recognition and segment selection. Chance accuracy is 20\% for environmental-sound recognition, 33.3\% for emotion recognition, and 50\% for gender recognition and segment selection. For environmental sounds, both recognition and selection decrease with increasing SNR in 15 of 16 model–task settings, indicating a shared dependence on cue salience as the environmental sound becomes weaker. For gender, recognition is generally strong, with Qwen2.5-Omni and MiniCPM-o approaching ceiling performance. Yet, this strong recognition does not always carry over to segment selection. For example, on math QA, Qwen2.5-Omni achieves 95.42\% gender-recognition accuracy but only 76.67\% segment-selection accuracy, suggesting that the remaining errors arise beyond atomic recognition and may reflect imperfect cue-to-segment binding. Emotion shows a different pattern. Recognition ranges from 35.50\% to 49.00\%, only modestly above chance, whereas selection ranges from 50.67\% to 73.00\%. On factual QA, Qwen2.5-Omni has lower emotion-recognition accuracy than Kimi-Audio (35.75\% vs.\ 49.00\%), yet achieves higher selection accuracy (73.00\% vs.\ 61.83\%), showing that explicit recognition accuracy does not directly determine cue-conditioned selection.


\textbf{Strong performance on atomic tasks does not consistently carry over to compositional tasks across models, cues and downstream tasks.} Table~\ref{tab:main_results} reports atomic and compositional downstream performance for each model. Using LLM-based parsing, we observe substantial degradation across nearly all model–task–cue settings. For example, on factual QA under E20, MiniCPM-o's ASR WER increases from 8.55\% to 67.97\%, while on math QA under E20, Qwen2.5-Omni's accuracy drops from 83.50\% to 40.50\%. The magnitude of degradation is also consistent with the selection results in Table~\ref{tab:attr_selection}. MiniCPM-o, whose gender selection is near ceiling, shows a much smaller drop on gender-conditioned QA than Qwen2.5-Omni. Environmental-sound conditions further reveal a trade-off between speech clarity and cue salience. As SNR increases, speech becomes easier to recognize while the environmental cue becomes weaker. For Qwen2.5-Omni and MiniCPM-o, compositional ASR WER therefore increases with SNR even as atomic WER decreases, showing that improved component performance does not necessarily translate to better compositional performance. Finally, the gap between tag-based and LLM-based parsing is generally larger for compositional than atomic tasks, with exceptions such as Kimi-Audio on factual QA. In sum, these results show that compositional degradation varies with selection behavior, cue salience, and output formatting. We analyze this format effect further in Sec.~\ref{sec:format}.

Finally, these results indicate that the compositionality gap does not arise at a single and common stage. Atomic recognition, segment selection, and downstream execution can each be good in isolation while their composition still degrades. Moreover, where the degradation concentrates differs across cue types and models, from cue-to-segment binding to output formatting. Therefore, diagnosing the compositionality gap requires measuring each stage separately rather than reading a single end-to-end score.

\subsection{Format Following}
\label{sec:format}

\begin{table}[t]
\centering
\caption{
Output-format compliance and parsing sensitivity. Fmt. denotes format-following rate (\%). For QA, \(\Delta\) is the accuracy difference between LLM-based and tag-based parsing; for ASR, it is the corresponding reduction in WER. Positive values indicate better scores with LLM-based parsing.
}
\label{tab:format_analysis}
\small
\resizebox{0.95\linewidth}{!}{
\begin{tabular}{llcccc}
\toprule
\textbf{Model} & \textbf{Stage}
& \textbf{Fmt.}
& \textbf{$\Delta_{\text{ASR}}$}
& \textbf{$\Delta_{\text{math}}$}
& \textbf{$\Delta_{\text{qa}}$} \\
\midrule

\multirow{2}{*}{Qwen2.5-Omni}
& Atomic
& 99.98
& -0.83
& +0.15
& +0.02 \\
& Comp.
& 100.00
& -1.66
& +9.87
& +0.36 \\
\midrule

\multirow{2}{*}{MiniCPM-o}
& Atomic
& 92.71
& +7.20
& +0.08
& +0.02 \\
& Comp.
& 54.91
& +12.67
& +48.87
& +8.57 \\
\midrule

\multirow{2}{*}{MiMo-Audio}
& Atomic
& 88.88
& +7.59
& +4.81
& +2.25 \\
& Comp.
& 83.43
& +1.08
& +3.73
& +3.73 \\
\midrule

\multirow{2}{*}{Kimi-Audio}
& Atomic
& 59.53
& +12.29
& +0.40
& +19.99 \\
& Comp.
& 56.27
& +13.96
& +13.07
& +12.61 \\

\bottomrule
\end{tabular}
}

\vspace{-15pt}
\end{table}


Table~\ref{tab:format_analysis} shows that composition can introduce additional output-format instability, although the effect is strongly model-dependent. Qwen2.5-Omni maintains near-perfect compliance across both stages, while MiMo-Audio shows only mild degradation. In contrast, MiniCPM-o drops from 92.71\% format compliance on atomic tasks to 54.91\% under composition, accompanied by a much larger difference between tag-based and LLM-based parsing. Kimi-Audio exhibits poor compliance already at the atomic stage, with similarly large parsing sensitivity throughout. These results indicate that, for some models, part of the measured atomic-to-compositional degradation is accompanied by a deterioration in output control rather than task performance alone.


Format compliance and parser sensitivity are related but not equivalent, since a large difference between the two parsing strategies only shows that the measured score depends strongly on output extraction rather than revealing the source of the discrepancy itself. This distinction is evident for Qwen2.5-Omni on compositional math QA, where the model often produces well-formed tags but places the full calculation inside the answer field, causing strict exact-match failures that LLM-based parsing can recover. We therefore report both parsing strategies throughout to capture not only task-level degradation under composition but also changes in output-format compliance and sensitivity to extraction.

\subsection{Positional Bias}
\label{sec:positionbias}

\begin{table}[t]
\centering
\caption{
Positional preference in segment selection. Entries show the percentage of predictions selecting the first segment, averaged over math and factual QA. With balanced target positions, 50\% indicates no positional preference.
}
\label{tab:position_bias}
\small
\resizebox{0.95\linewidth}{!}{
\begin{tabular}{lccccc}
\toprule
\textbf{Model}
& \textbf{E0}
& \textbf{E10}
& \textbf{E20}
& \textbf{Gender}
& \textbf{Emotion} \\
\midrule

Qwen2.5-Omni & 54.67 & 55.42 & 56.25 & 63.42 & 52.50 \\
MiniCPM-o    & 41.00 & 35.34 & 29.08 & 44.34 & 27.09 \\
MiMo-Audio   & 75.83 & 79.17 & 83.07 & 52.17 & 84.50 \\
Kimi-Audio   & 43.50 & 38.25 & 34.42 & 47.34 & 41.42 \\

\bottomrule
\end{tabular}
}

\vspace{-15pt}
\end{table}

Sec.~\ref{sec:main_results} identifies cue-conditioned segment selection as a fragile stage in compositional tasks. To examine how selection fails, Table~\ref{tab:position_bias} reports the fraction of responses selecting the first segment for each model under each cue type, where 50\% indicates no positional preference. The models split into two groups, Qwen2.5-Omni and MiMo-Audio favoring the first segment and MiniCPM-o and Kimi-Audio favoring the second.



The strength of these positional preferences varies substantially across cue types. For gender and emotion, the magnitude is model-dependent, with some cases close to the 50\% baseline and others showing strong preferences. Environmental sounds show a more systematic pattern. For all four models, the deviation from 50\% increases monotonically from E0 to E20 as the cue becomes less salient. MiniCPM-o, for example, shifts from 41.00\% at E0 to 29.08\% at E20. A fixed positional preference would produce similar deviations across conditions. Instead, the preference strengthens as the environmental cue weakens, suggesting that positional tendencies become more pronounced when cue-based selection is less reliable. Unlike the answer-option order bias reported in LALMs~\cite{hearingorder}, the positional preference observed here is cue-dependent rather than condition-invariant.

\section{Can Explicit Reasoning Mitigate the compositionality gap?}
\label{sec:cot}

\begin{table}[t]
\centering
\caption{
Effect of CoT prompting on compositional QA. Values are changes from direct prompting, averaged across cue conditions; \(\Delta_{\mathrm{Tag}}\), \(\Delta_{\mathrm{LLM}}\), and \(\Delta_{\mathrm{Fmt}}\) denote changes in tag-parsed accuracy, LLM-parsed accuracy, and format-following rate, all in percentage points.
}
\label{tab:cot_analysis}
\small
\resizebox{0.95\linewidth}{!}{
\begin{tabular}{lrrr|rrr}
\toprule
& \multicolumn{3}{c|}{\textbf{Math QA}}
& \multicolumn{3}{c}{\textbf{Factual QA}} \\
\cmidrule(lr){2-4}
\cmidrule(lr){5-7}

\textbf{Model}
& $\Delta_{\mathrm{Tag}}$
& $\Delta_{\mathrm{LLM}}$
& $\Delta_{\mathrm{Fmt}}$
& $\Delta_{\mathrm{Tag}}$
& $\Delta_{\mathrm{LLM}}$
& $\Delta_{\mathrm{Fmt}}$ \\
\midrule

Qwen2.5-Omni & -1.56 & -11.17 & 0.00 & +2.10 & +1.90 & -0.03 \\
MiniCPM-o    & +41.67 & -2.10 & +68.50 & +10.10 & +14.63 & -12.07 \\
MiMo-Audio   & +24.06 & +20.53 & +9.50 & +10.53 & +9.54 & +25.43 \\
Kimi-Audio   & -4.37 & +6.10 & -15.17 & +3.27 & +12.71 & +27.97 \\

\bottomrule
\end{tabular}
}
\vspace{-15pt}
\end{table}


To examine whether explicitly structuring the compositional process can improve performance, we evaluate chain-of-thought (CoT) prompting~\cite{cot,visualcot} on compositional math and factual QA. The prompt makes the intended execution order explicit within a single inference (identify the cue-matched segment, recover its question, and answer it), while requiring only the final answer to appear inside the same tags used by the baseline. All evaluation metrics and parsing procedures remain unchanged.


Table~\ref{tab:cot_analysis} reports changes in accuracy and format compliance under CoT prompting. Its effect on answer quality varies across models and tasks. On math QA, LLM-parsed accuracy decreases for Qwen2.5-Omni and MiniCPM-o but improves for MiMo-Audio and Kimi-Audio. In contrast, all four models improve on factual QA, with gains ranging from 1.90 to 14.63 percentage points. Answer quality does not consistently track format compliance. For example, MiniCPM-o gains 41.67 percentage points in tag-parsed math accuracy, largely alongside a 68.50-point increase in format compliance, while its LLM-parsed accuracy changes little. Conversely, Kimi-Audio improves by 6.10 percentage points in LLM-parsed math accuracy despite a 15.17-point decline in compliance. Qwen2.5-Omni shows that CoT can harm answer quality, losing 11.17 percentage points in LLM-parsed math accuracy while compliance remains unchanged. In summary, CoT does not consistently reduce the compositionality gap, and its effects depend on both the downstream task and the source of performance change.

\section{Conclusion}
\label{sec:conclusion}


We conduct a controlled diagnostic study of capability composition in LALMs. By separately measuring component capabilities, cue-conditioned segment selection, and compositional downstream performance, we examine where failures arise when individually available capabilities must be combined. Across models, acoustic cues, and downstream tasks, we observe a consistent degradation from atomic to compositional settings. The relationship between atomic recognition and segment selection varies across cue types, while cue-dependent positional tendencies further show that selection behavior cannot be explained by recognition accuracy alone. We further find that format compliance degrades under composition but does not fully explain the observed degradation, while step-by-step CoT prompting does not provide a general solution for the compositionality gap. This gap may limit the reliability of LALMs in more complex audio tasks, where relevant content must first be identified from acoustic cues before downstream processing can be performed.

\section{Acknowledgement}

We would like to thank Chung-Cheng Chen, Jen-Hao Cheng, Tsung-Ying Yang, Hsiao-Tsung Hung and Dau-Cheng Lyu at ASUS-OCIS for their helpful and inspiring discussions and feedback throughout this work. We are also grateful to Hung-Ting Su for his valuable comments and suggestions on the manuscript.

\bibliographystyle{IEEEbib}
\bibliography{refs}

\end{document}